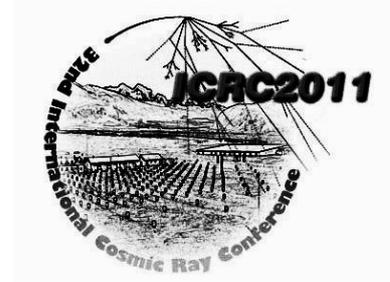

# Overview of the results from extra-galactic observations with the MAGIC telescopes

KARSTEN BERGER[1], ON BEHALF OF THE MAGIC COLLABORATION[2]
[1]*Instituto de Astrofisica de Canarias and Universidad de La Laguna, La Laguna, Spain*
[2]*The full collaborator list can be found at: wwwmagic.mppmu.mpg.de*
berger.karsten@gmail.com

**Abstract:** The non-thermal jet emission in active galactic nuclei covers several orders of magnitude in the frequency range. Hence the observational approach needs multi-wavelength (MWL) campaigns collecting data in the radio, optical, UV, X-rays, high energy until the Very High Energy (VHE) $\gamma$-ray band. MAGIC, a system of two 17 m diameter telescopes at the Roque de los Muchachos observatory on the canary island La Palma, actively participates and organizes MWL observations on known and newly discovered VHE sources. In these proceedings we report the latest results of extra-galactic observations with MAGIC, which gained new insights in time variability studies and jet emission models.

## 1 The MAGIC telescopes

The MAGIC telescopes are situated on the Roque de los Muchachos on the canary island La Palma, 28°N, 18°W at a height of 2200 m. They are separated by 85 m and work in a coincidence trigger mode. Thanks to the 3D shower reconstruction a two times better integral sensitivity above 250 GeV (better than 0.8% of the Crab Nebula flux in 50 h of observation) is achieved. Differentially the sensitivity improved especially below 100 GeV, where a lower background rate (factor ~10) significantly reduces the systematic errors. The angular resolution (defined as the sigma of a 2D Gaussian) is 0.07° on an event to event basis. The energy resolution (above 300 GeV) improved to 15%. More details can be found in a dedicated contribution to this conference [1].

## 2 New results from 3C 279

3C 279 is the most distant VHE $\gamma$-ray emitter detected so far, at a redshift of 0.536 [2]. After the initial discovery of a short flare in February 2006 with the stand-alone MAGIC-I telescope an observation during a historical optical outburst in January 2007 resulted in the second detection on 16th of the same month at a confidence level of 5.4$\sigma$ [3]. The measured spectrum is compatible with the hard spectrum of the 2006 campaign.

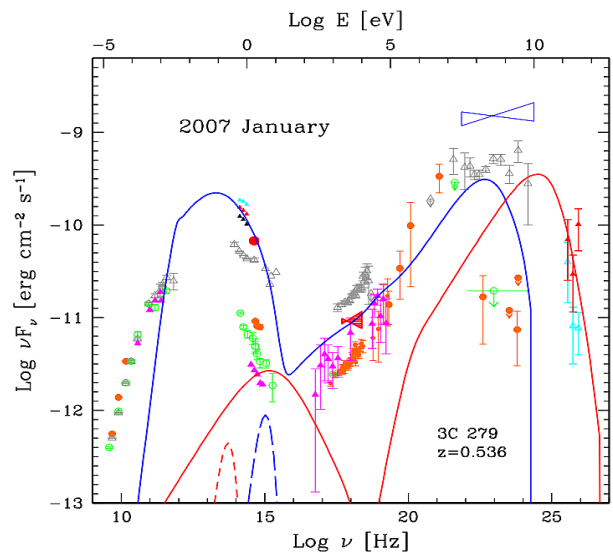

Figure 1. SED of 3C 279 during the 2007 observation campaign. The blue line corresponds to a model fit of the emission from the zone inside the BLR while the red line denotes the emission from the second zone outside the BLR. A more detailed description of the model, the used dataset and other model fits can be found in [3].

The VHE flare is not accompanied by a distinct feature in the optical or X-ray wavebands [3], however the electron vector polarization angle rotates and a new knot in the jet emerges [4]. A detailed discussion of the results of the modeling of the spectral energy distribution (SED) is beyond the scope of this proceeding. It is however important to note that a simple one zone synchroton self Compton plus external Compton (SSC+EC) model with an emission region inside the broad line region (BLR) cannot reproduce our data [3]. Instead we used a two zone model (see Fig. 1) and a lepto-hadronic model. Both can describe the data satisfactorily; however they also include a larger number of free parameters. The new observations cannot be used to constrain the extragalactic background light (EBL) significantly, due to the maturity of the EBL models, which imply a highly transparent Universe to VHE photons [5]. The 2009 observations did not yield a detection, which is compatible with the extrapolation of the soft Fermi spectrum at lower energies. Several questions remain: What is the exact variability time scale of the VHE flares? Is there a delay with respect to the GeV γ-ray emission? 3C 279 is thus an interesting target for future observations with the stereoscopic MAGIC telescope system.

## 3  The BL Lac object B3 2247+381

B3 2247 (z=0.119) is another example of a successful optical trigger from the Tuorla monitoring program[1]. A more detailed discussion about the use of optical triggers by MAGIC can be found in [6]. The 14h MAGIC observation resulted in a clear detection at a level of 2.3% C.U. (above 200 GeV), which is consistent with the previously reported upper limit [7]. We can thus not conclude that the source was in a VHE γ-ray high state. However, both the optical and the X-ray state show a clear enhancement compared to previous observations [8]. We can model the SED with a simple one zone SSC model and derive parameters that are typical for HBLs (see Fig. 2).

## 4  Summary of five observation years from PG 1553+113

The redshift of PG 1553 is still unkown, however recent estimates indicate that it is between 0.4 and 0.45 [9]. The source was discovered in VHE γ-rays in 2005 by H.E.S.S. [10] and MAGIC [11]. Follow-up observations over five years have resulted in a very rich dataset that is accompanied by optical (R-band), X-ray and (since June 2008) also high energy γ-ray data (Fig. 3). The source is continuously detected, with significant variability between 4% and 11% C.U. above 150 GeV. We also found a mild correlation between the VHE γ-ray and the optical R-band flux [12]. The spectral characteristics are stable over the years with a mean power law photon index of -4.27±0.14. The SED shows two narrow peaks and an excellent agreement between Fermi and MAGIC in the overlapping energy region. The SSC model parameters are typical for hard spectrum HBLs and imply a relatively large minimum Lorentz factor ($2.5 \cdot 10^3$) of the electron population [12].

---

[1] http://users.utu.fi/kani/1m/index.html

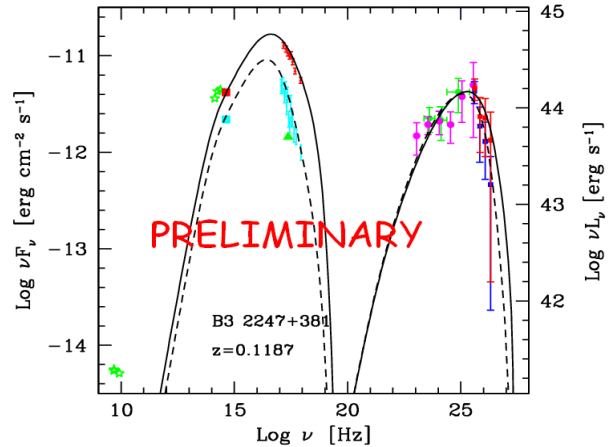

Figure 2. SED of B3 2247+381 from radio to VHE γ-rays. Green and light blue points represent non-simultaneous low state data. The solid (dotted) line is our SSC-model fit to the high (low) state observations.

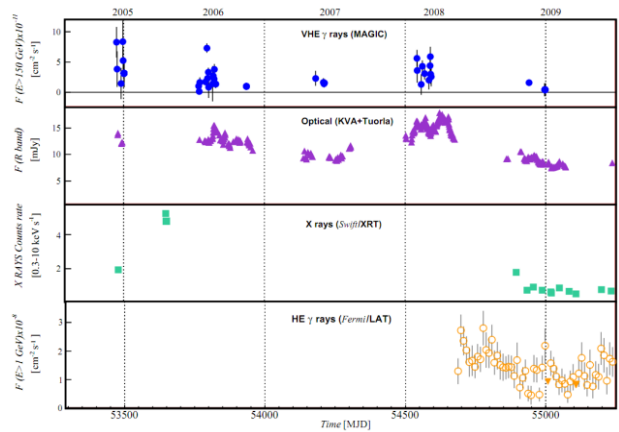

Figure 3. Light curve of PG 1553+113 in VHE γ-rays, optical, X-rays and HE γ-rays over five observation years [12].

## 5  1ES 0806+524 in a flaring state

Though 1ES 0806, a high peaked BL Lac object (z=0.138), was classified as potential TeV γ-ray emitter by [13], the first observations with the stand-alone MAGIC-I telescope resulted only in upper limits [7]. The source was subsequently discovered by VERITAS [14] at a flux level of 1.8% C.U. (E>300 GeV), an order of magnitude below the prediction. The weakness of the source left the question of variability unclear and resulted in a spectrum with large errors in a narrow energy range between 300 to 700 GeV. In October 2010 the source started to show enhanced optical activity which finally triggered observations with both MAGIC telescopes in February 2011. Most noteworthy is a VHE γ-ray flare on February 24$^{th}$ at a level of 5.6% C.U. resulting in a ~6.7σ detection in just 3.3h [15]. A comparable observation time on the day before did not detect the source, which strongly suggests day-scale variability. Follow up observations with Swift revealed enhanced high X-ray emission (a factor 2-3 above average, see [16]). The spectral and variability analysis of the full 20h dataset is underway.

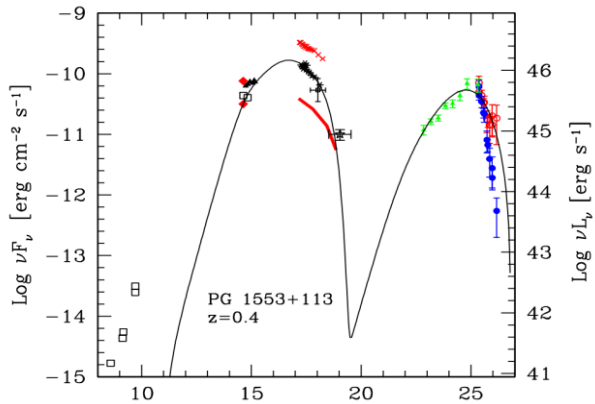

Figure 4. Average SED for PG 1553+113 during five years of observations [12] with an SSC model fit (black line).

## 6  Swift J164449.3+573451

The GRB nature of this Swift trigger was rapidly ruled out due to the unusually long lasting flaring activity. MAGIC started observations on 2011/03/31 at 02:22 UT (~2.5 days after the trigger). We have collected 27h of good quality data during 12 nights with an energy threshold of about 150 GeV. Neither the full dataset, nor data taken during high X-ray fluxes ($>10^{-10}$ erg cm$^{-2}$ s$^{-1}$) shows a significant excess. We have thus calculated preliminary upper limits (see Fig. 5).

## 7  PKS 1424+240

This intermediate peaked BL Lac object with unkown redshift has been discovered by Veritas in 2009 [17]. MAGIC has observed the source since 2006 using first only MAGIC-I and starting from 2010 also the second telescope. After upper limits in 2006 and 2007 the source was finally detected in 2009 during enhanced optical emission and has been detected ever since (see Fig. 6). The flux varies on yearly time scales (note the enhanced flux state in 2011), while the spectrum appears to be stable within errors and can be described by a simple power law (Fig. 7). The modeling of the spectra energy distribution together with data from the Fermi satellite is in progress.

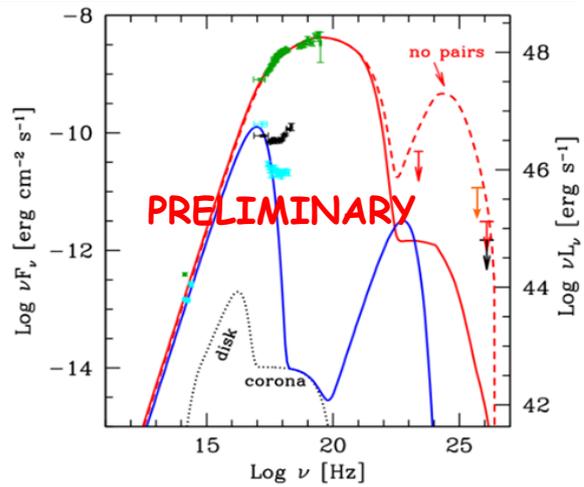

Figure 5. SED of Swift J164449.3+573451 reproduced from [18]. The preliminary MAGIC upper limits are shown as orange arrow.

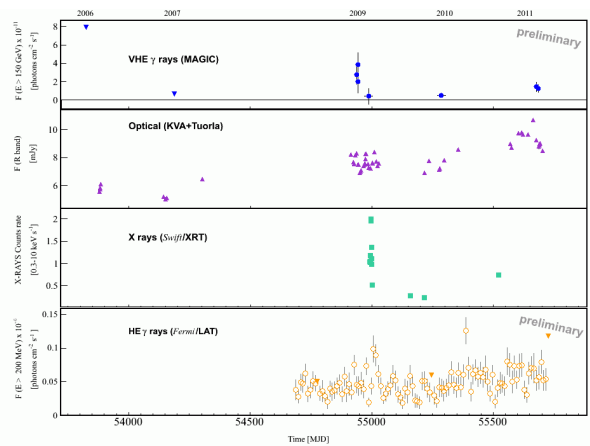

Figure 6. Light curve of PKS 1424+240 in the VHE and HE γ-ray as well as the optical and X-ray band.

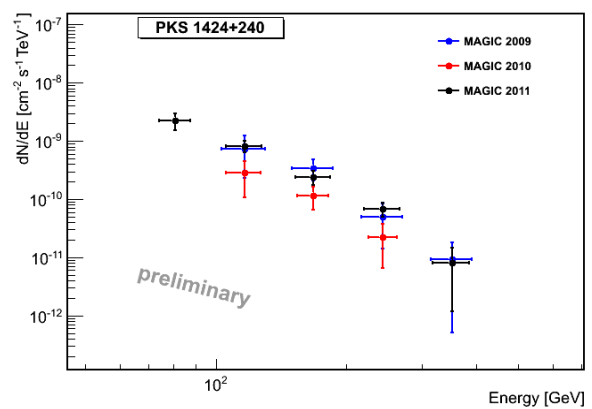

Figure 7. Differential energy spectrum of PKS 1424+240 as observed by MAGIC in the years 2009, 2010 and 2011.

## 8 A new discovery: 1ES 1741+196

The host galaxy of 1ES 174+196 (z=0.083) is one of the largest and most luminous of all BL Lacs. The authors of [19] have found two more galaxies that seem to form a gravitationally bound triplet in the 1ES 1741 system, potentially connected with tidal streams. MAGIC observed 1ES 1741 for more than 60 h with the new stereoscopic setup and has discovered (at the 5.1σ confidence level) VHE γ-rays above 250 GeV from the system (see Fig. 10). The integral flux corresponds to 0.8% C.U., which makes 1ES 1741 one of the weakest AGNs detected so far and the first source <1% C.U. detected by MAGIC. The measured flux is consistent with previously published upper limits [20].

The analysis is currently ongoing. The light curve and spectrum as well as the interpretation of the broad band SED will be published in a dedicated article.

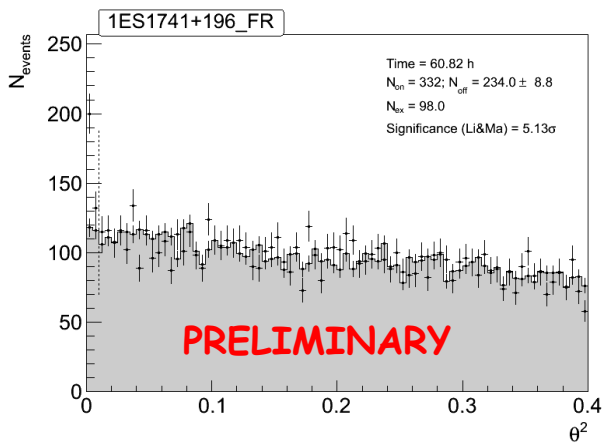

Figure 8. Theta squared distribution for the MAGIC observations of 1ES1741+196. Observation details are given in the inlay of the figure. The analysis includes events above 250 GeV.

## 9 Conclusions

The MAGIC stereo system is working extremely well, actively extending our knowledge of the extra-galactic VHE γ-ray universe. Compared to the monoscopic observations the discovery frequency of AGNs in VHE γ-rays has significantly increased, thanks to the improved sensitivity, the high energy γ-ray catalogue provided by the Fermi satellite and optical triggers issued by the Tuorla blazar monitoring program. We are actively participating on large scale multi-wavelengths campaigns that help to constrain theoretical model predictions. After a planned camera and readout upgrade during summer 2011, MAGIC will further extend its capabilities.

## 10 Acknowledgements

We would like to thank the Instituto de Astrofisica de Canarias for the excellent working conditions at the Observatorio del Roque de los Muchachos in La Palma. The support of the German BMBF and MPG, the Italian INFN, the Swiss National Fund SNF, and the Spanish MICINN is gratefully acknowledged. This work was also supported by the Marie Curie program, by the CPAN CSD2007-00042 and MultiDark CSD2009-00064 projects of the Spanish Consolider-Ingenio 2010 programme, by grant DO02-353 of the Bulgarian NSF, by grant 127740 of the Academy of Finland, by the YIP of the Helmholtz Gemeinschaft, by the DFG Cluster of Excellence "Origin and Structure of the Universe", by the DFG Collaborative Research Centers SFB823/C4 and SFB876/C3, and by the Polish MNiSzW grant 745/N-HESS-MAGIC/2010/0.